\documentclass[conference]{IEEEtran}
\IEEEoverridecommandlockouts

\usepackage{cite}
\usepackage{amsmath,amssymb,amsfonts}
\usepackage{algorithmic}
\usepackage{graphicx}
\usepackage{textcomp}
\usepackage{xcolor}

\usepackage{cite}
\usepackage{amsmath,amssymb,amsfonts}
\usepackage{algorithmic}
\usepackage{graphicx}
\usepackage{textcomp}
\usepackage{xcolor}
\usepackage[hyphens]{url}
\usepackage{fancyhdr}
\usepackage[bookmarks=true,breaklinks=true,letterpaper=true,colorlinks,citecolor=blue,linkcolor=blue,urlcolor=blue]{hyperref}

\usepackage{color}
\usepackage{soul}
\usepackage{multirow}
\usepackage{makecell, rotating}
\usepackage{subfigure}
\usepackage{enumerate}
\usepackage{enumitem}
\usepackage{balance}
\usepackage[linesnumbered,ruled,lined]{algorithm2e}
\usepackage{threeparttable}
\usepackage{caption}
\usepackage{amsmath}
\usepackage{amssymb}
\usepackage{algorithmic}
\usepackage{graphicx}
\usepackage{float}
\usepackage{rotating}
\usepackage{booktabs}

\newcommand{\RebuttalChange}[1]{\textcolor{black}{#1}}

\newcommand{\SolutionName}{Trinity}
\newcommand{\Tofill}[1]{\textcolor{black}{#1}}

\def\BibTeX{{\rm B\kern-.05em{\sc i\kern-.025em b}\kern-.08em
    T\kern-.1667em\lower.7ex\hbox{E}\kern-.125emX}}
\begin{document}

\title{{\SolutionName}: A General Purpose FHE Accelerator
\thanks{Xianglong Deng and Shengyu Fan contributed equally to this work. The corresponding author is Mingzhe Zhang (mingzhe-zhang@outlook.com).}
}

\DeclareRobustCommand*{\IEEEauthorrefmark}[1]{%
    \raisebox{0pt}[0pt][0pt]{\textsuperscript{\footnotesize\ensuremath{#1}}}}

\author{\IEEEauthorblockN{Xianglong Deng\IEEEauthorrefmark{1,2}, Shengyu Fan\IEEEauthorrefmark{1,2}, Zhicheng Hu\IEEEauthorrefmark{3}, Zhuoyu Tian\IEEEauthorrefmark{1,2}, Zihao Yang\IEEEauthorrefmark{1,2}, Jiangrui Yu\IEEEauthorrefmark{4,5},}
\IEEEauthorblockN{Dingyuan Cao\IEEEauthorrefmark{6}, Dan Meng\IEEEauthorrefmark{1}, Rui Hou\IEEEauthorrefmark{1}, Meng Li\IEEEauthorrefmark{4,5}, Qian Lou\IEEEauthorrefmark{7} and Mingzhe Zhang\IEEEauthorrefmark{1}}
\IEEEauthorblockA{\IEEEauthorrefmark{1}\textit{Key Laboratory of Cyberspace Security Defense, Institute of Information Engineering, CAS, Beijing, China}}
\IEEEauthorblockA{\IEEEauthorrefmark{2}\textit{School of Cyber Security, University of Chinese Academy of Sciences, Beijing, China}}
\IEEEauthorblockA{\IEEEauthorrefmark{3}\textit{University of Electronic Science and Technology of China, Chengdu, China}}
\IEEEauthorblockA{\IEEEauthorrefmark{4}\textit{Institute for Artificial Intelligence, Peking University, Beijing, China}}
\IEEEauthorblockA{\IEEEauthorrefmark{5}\textit{School of Integrated Circuits, Peking University, Beijing, China}}
\IEEEauthorblockA{\IEEEauthorrefmark{6}\textit{University of Illinois, Urbana-Champaign, IL, USA}}
\IEEEauthorblockA{\IEEEauthorrefmark{7}\textit{University of Central Florida, Orlando, FL, USA}}
\IEEEauthorblockA{\texttt{\textit{\{dengxianglong, fanshengyu, tianzhuoyu, yangzihao, mengdan, hourui\}@iie.ac.cn}}}
\IEEEauthorblockA{\texttt{\textit{zhichenghu@std.uestc.edu.cn}}, 
\texttt{\textit{meng.li@pku.edu.cn}},
\texttt{\textit{jiangrui.yu@stu.pku.edu.cn}}}
\IEEEauthorblockA{  \texttt{\textit{dc29@illinois.edu}}, \texttt{\textit{qian.lou@ucf.edu}}, \texttt{\textit{mingzhe-zhang@outlook.com}}
}
}

\maketitle

\begin{abstract}
Fully Homomorphic Encryption (FHE) is crucial for privacy-preserving computing, which allows direct computation on encrypted data. While various FHE schemes have been proposed, none of them efficiently support both arithmetic FHE and logic FHE simultaneously. To address this issue, researchers explore the combination of different FHE schemes within a single application and propose algorithms for the conversion between them. 
Unfortunately, all prior ASIC-based FHE accelerators are designed to support a single FHE scheme, and none of them supports the acceleration for FHE scheme conversion. This necessitates FHE acceleration systems to integrate multiple accelerators for different schemes, leading to increased system complexity and hindering performance enhancement.


In this paper, we present the first multi-modal FHE accelerator based on a unified architecture, which efficiently supports CKKS, TFHE, and their conversion scheme within a single accelerator. To achieve this goal, we first analyze the theoretical foundations of the aforementioned schemes and highlight their composition from a finite number of arithmetic kernels. Then, we investigate the challenges for efficiently supporting these kernels within a unified architecture, which include 1) concurrent support for NTT and FFT, 2) maintaining high hardware utilization across various polynomial lengths, and 3) ensuring consistent performance across diverse arithmetic kernels. To tackle these challenges, we propose a novel FHE accelerator named {\SolutionName}, which incorporates algorithm optimizations, hardware component reuse, and dynamic workload scheduling to enhance the acceleration of CKKS, TFHE, and their conversion scheme. \RebuttalChange{By adaptive select the proper allocation of components for NTT and MAC, {\SolutionName} maintains high utilization across NTTs with various polynomial lengths and imbalanced arithmetic workloads.} {The experiment results show that, for the pure CKKS and TFHE workloads, the performance of our {\SolutionName} outperforms the state-of-the-art accelerator for CKKS (SHARP) and TFHE (Morphling) by \RebuttalChange{1.49$\times$} and {4.23$\times$}, respectively. }Moreover, {\SolutionName} achieves \Tofill{919.3$\times$} performance improvement for the FHE-conversion scheme over the CPU-based implementation. 
Notably, despite the performance improvement, the hardware overhead of {\SolutionName} is only \Tofill{85\%} of the summed circuit areas of SHARP and Morphling.

\end{abstract}

\begin{IEEEkeywords}
CKKS, TFHE, Scheme Conversion, ASIC, FHE.
\end{IEEEkeywords}

\section{Introduction}

Fully Homomorphic Encryption (FHE) is a novel encryption method that enables direct computations on encrypted data. This capability allows users to perform secure computation on private data in untrusted environments, making FHE a valuable technique for privacy-preserving computing.

Several FHE schemes have been proposed, which can be categorized into two types: the arithmetic FHE (i.e. BGV\cite{brakerski2014bgv}, BFV\cite{brakerski2012bfv}, and CKKS\cite{cheon2017ckks}) and logic FHE (i.e. FHEW\cite{ducas2015fhew} and TFHE\cite{chillotti2020tfhe}). 
Arithmetic FHE allows SIMD-style arithmetic operations by packing multiple plaintexts into a ciphertext. In contrast, logic FHE encrypts a single bit into a ciphertext and allows logic operations such as comparison. 
The real-world applications contain both arithmetic and logic operations, however, none of the FHE schemes can support both of them. Therefore, researchers have started to study the construction of hybrid-scheme applications based on different types of FHE schemes and have proposed several Scheme Conversion algorithms\cite{chen2021homoconversion, lu2021pegasus, al2022openfhe} between different types of FHE schemes.



Hybrid-scheme applications pose challenges to the FHE accelerator. Several FHE accelerators\cite{agrawal2023fab, kim2022ark, kim2023sharp, samardzic2021f1, samardzic2022craterlake} have been proposed, but they only provide efficient support for one type of FHE scheme. For example, SHARP\cite{kim2023sharp} can efficiently support the 36-bit version of the CKKS scheme, which outperforms CPU by \Tofill{22939$\times$} with reasonable hardware overheads. Morphling improves the TFHE workloads over CPU by \Tofill{2343$\times$}. Furthermore, to the best of our knowledge, there is no FHE accelerators that can support the Scheme Conversion algorithm. Therefore, a general FHE system should be implemented with heterogeneous accelerators, which increases the system complexity and probably decrease the overall performance.

In this paper, we make a comprehensive investigation for the CKKS, TFHE, and the conversion schemes, and then present several key observations: 
1) both schemes and the conversion algorithm consist of a finite set of kernels; 
2) the Fast Fourier Transform (FFT) used in TFHE can be approximately replaced by the Number Theoretic Transform (NTT); 
3) the key to uniform hardware design is maintaining high utilization across various workloads. 
Based on the above observations, we propose an accelerator named \emph{\SolutionName}, which efficiently supports the above schemes within the unified architecture as follows:
1) We substitute FFT with NTT in TFHE by selecting an appropriate modulus $q$;
2) We design a configurable component that dynamically supports NTT across different polynomial lengths;
3) We implement a component reuse strategy that employs some MAC units for NTT computations, ensuring workload balance across various workloads.


We evaluate our proposed \emph{\SolutionName} within both CKKS and TFHE applications. The experimental results show that, within CKKS applications, \emph{\SolutionName} achieves an average speedup over SHARP\cite{kim2023sharp} by \RebuttalChange{1.49$\times$}. For TFHE applications, \emph{\SolutionName} outperforms Morphling\cite{prasetiyo2024morphling} by \Tofill{4.23$\times$}. When considering the conversion between CKKS and TFHE, \emph{\SolutionName} outperforms CPU-based implementation by \Tofill{919.3$\times$}. When considering the hardware overhead, the area of \emph{\SolutionName} is smaller than the total area of SHARP and Morphling by \Tofill{15\%}. 


The contributions of this work are as follows:

\begin{itemize}
    \item For the first time, we make detailed investigation for the opportunities and challenges of efficiently support various FHE schemes and their conversion algorithms within the unified architecture. These experiences bridge the cryptography and computer architecture, which will probably inspire the future works.
    \item We explore to provide high-performance support for various FHE kernels (i.e., NTT with different polynomial lengths, and NTT and MAC workload with changing proportion) within the heterogeneous architecture and dynamic schedule.
    \item We experimentally prove that, it is possible to architect high-performance multi-modal accelerator based on reasonable hardware overhead.
\end{itemize}

\section{Background}
\begin{table}[]
    \centering
    \caption{Notations, parameters and operations. }
    \begin{tabular}{l|l}
    \toprule[1pt]
    \textbf{Notation}     & \textbf{Description} \\
    \midrule[0.5pt]
    \textbf{m}     & Plaintext vector.  \\
    $[[\textbf{m}]]$     & \makecell[tl]{RLWE ciphertext of \textbf{m}.  $\left[\left[\textbf{m} \right]\right]$= (a(X), b(X)).} \\
    m     & Scalar plaintext.  \\
    $[[m]]$     & \makecell[tl]{LWE ciphertext of scalar message {m}.  $\left[\left[{m} \right]\right]$= (\textbf{a}, b).} \\
    $P(X)$ & Polynomial. \\
    $P_\textbf{m}$ & Plaintext polynomial corresponding to message \textbf{m}. \\
    n & The number of slots in a ciphertext. \\
    N & The polynomial size. \\
    $\Delta$ & The scale factor. \\
    $R_{Q}$ & The polynomial ring, $R_Q = Z_Q / (X^{N} + 1)$. \\
    $Q$ & The polynomial modulus $Q = \prod_{i=0}^{L} q_i$. \\
    $P$ & The special prime $P = \prod_{i=0}^{\alpha} p_i$. \\
    $q_i$ & The small RNS moduli composing $Q$. \\
    $p_i$ & The small RNS moduli composing $P$. \\
    $L$ & The maximum level of polynomial. \\
    $l$ & The current level of polynomial. \\
    $R_{q_l}$ & Polynomial at level l. \\
    $\text{dnum}$ & The decomposition number. \\
    $\alpha$ & \makecell[lt]{The number of RNS moduli in a digit. $\alpha = \lceil \frac{L+1}{dnum} \rceil$. }\\
    $\beta$ & $\beta = \lceil \frac{l+1}{dnum} \rceil$. \\
    $\text{evk}$ & The evaluation key. \\
    $\text{n}_{\text{lwe}}$ & The dimension of LWE ciphertext. \\
    $k$ & The dimension of GLWE ciphertext. \\
    $q$ & The modulus of coefficients. \\
    $l_b$ & The decomposition level of bootstrapping key. \\
    $l_k$ & The decomposition level of TFHE keyswitching key. \\
    $\text{ACC}_i$ & The accumulation ciphertext in i-th iteration. \\
    $\text{ksk}$ & The keyswitch key in TFHE. \\
    $\text{bsk}$ & The bootstrapping key in TFHE. \\
    $\text{n}_{\text{slot}}$ & The number of valid slots in the RLWE ciphertext. \\
    
    $\text{Rotate}$ & Rotate([[\textbf{m}]]) $\to$ $(a(X) \cdot X^r, b(X) \cdot X^r)$.\\
    
    $\text{SampleExtract}$ & SampleExtract([[\textbf{m}]], i) $\to$ [[\textbf{m}[i]]] \\
    $\text{Decompose}$ & Decompose($P(X)$, $l$) $\to$ ($\tilde{P}(X)_{0}$,...,$\tilde{P}(X)_{l-1}$). \\
    \bottomrule[1pt]
    \end{tabular}
    \label{tab:background:notation}
\end{table}

\begin{table}[t]

    \centering
    \caption{Hierarchical reconstruction model of CKKS.}\label{tab:background:hierarchy}
    \begin{tabular}{l|l|l}\hline
    \toprule[1pt]
       \multirow{2}{*}{\textbf{Operation}} & \multirow{2}{*}{\textbf{Description}} & \textbf{Composing} \\ 
        && \textbf{Kernels} \\ 
        \midrule[0.5pt]
        
          HMult & {Multiply two ciphertexts} & \makecell[tl]{{NTT, BConv, IP,}\\{ModMul, ModAdd}}\\  
          PMult & {Multiply a ciphertext by a plaintext} & \makecell[tl]{{ModMul, ModAdd}}\\  
          HRotate & {Homomorphic rotation of ciphertext} & \makecell[tl]{{NTT, BConv, IP,}\\{ModMul, ModAdd,}\\{Auto}}\\  
          HAdd & {Add two ciphertexts} & \makecell[tl]{ModAdd}\\  
          PAdd & {Add a ciphertext and a plaintext} & \makecell[tl]{ModAdd}\\  
          Rescale & {Reduce the level of a ciphertext} & \makecell[tl]{NTT, ModAdd}\\

      
      \bottomrule[1pt]
    \end{tabular}
    \vspace{-15pt}
\end{table}

\begin{algorithm}[t]
	\caption{\texttt{Hybrid KeySwitch} ($[d]_{C_l}$, evk)} 
	\label{alg:background:ckks_keyswitch} 
	\begin{algorithmic} [1]
        \REQUIRE{$[d]_{C_l}$: a polynomial with level $l$, evk: the evaluation key}
        \STATE $[d']_{C'_{i}}$ = Decompose($[d]_{C_l}$, $\beta$) \\
        \STATE $\tilde{ct}_{1}$ = 0; $\tilde{ct}_{2}$ = 0\\
        \FOR{i = 0; i $<$ $\beta$; i = i + 1}
        \STATE $[\tilde{d}]_{D_{\beta}}$ = BConv$_{C'_{i} \to D_{\beta}}$($d'$[i])\\
        \STATE $[\tilde{d}]_{D_{\beta}}$ = NTT($[\tilde{d}]_{D_{\beta}}$)\\
        \ENDFOR
        \FOR{j = 0; j $<$ 2; j = j + 1}
            \FOR{i = 0; i $<$ $\beta$; i = i + 1}
            \STATE $[\tilde{ct}_{j}]_{D_{\beta}}$ = $[\tilde{ct}_{j}]_{D_{\beta}}$ + $[\tilde{d}]_{D_{\beta}}$ * evk$_{i,j}$
            \ENDFOR
            \STATE $[\tilde{ct}_{j}]_{D_{\beta}}$ = iNTT($[\tilde{ct}_{j}]_{D_{\beta}}$)\\
            \STATE $ct_{mult_{j}}$ = $[\tilde{ct}_{j}]_{C_l}$ - BConv$_{B \to C_{l}}$($[\tilde{ct}_{j}]_{B}$)\\
        \ENDFOR
        \RETURN $ct_{mult_{1}}$, $ct_{mult_{2}}$
        
	\end{algorithmic}  
\end{algorithm}



\label{sec:background:tfhe-ckks}

In this section, we first describe the data types, kernels, and operations utilized in both the CKKS and TFHE cryptographic schemes. Furthermore, we introduce the algorithm for the {Scheme Conversion between CKKS and TFHE}. To facilitate comprehension, we summarize all notations, parameters, and operations discussed in this paper in Table \ref{tab:background:notation}. 
Additionally, we summarize the hierarchical operation reconstruction model of CKKS in Table \ref{tab:background:hierarchy}. 

\subsection{CKKS scheme}


\label{sec:background:ckks_kernels}
CKKS is an arithmetic FHE, which enables SIMD-style arithmetic operations by packing multiple plaintexts into a ciphertext. The homomorphic operation of CKKS can be decomposed into the following kernels:

\begin{itemize}
    \item \textbf{NTT.} NTT accelerates polynomial multiplication, by transforming a polynomial from the coefficient representation to the evaluation representation. 
    \item \textbf{BConv.} BConv adjusts the coefficient modulus of a polynomial. The process involves multiplying a polynomial matrix of size ${\alpha} \times N$ by a base matrix of size $l \times {\alpha}$, effectively changing the modulus from $q_{\alpha}$ to $q_l$\cite{kim2022ark}. 
    \item \textbf{IP.} IP multiplies a polynomial with the evaluation key (evk) during KeySwitch. 
    IP operations can also be structured as a multiplication between a vector $[\tilde{d}_{0}, \dots, \tilde{d}_{dnum-1}]$ and an evk matrix of size $dnum \times 2$
    \item \textbf{ModMul. } ModMul performs the element-wise modular multiplication of two polynomials that are in evaluation representation.
    \item \textbf{ModAdd. } ModAdd performs the element-wise modular addition of two polynomials.
    \item \textbf{Auto. } Auto performs automorphism transform on a polynomial, which maps the indices of each coefficient from $i$ to $\sigma_{r}(i)$ ($\sigma_r(i) = i \cdot 5^r \mod N$).
\end{itemize}

Based on the above kernels, CKKS provides several homomorphic operations, which are shown in Table \ref{tab:background:hierarchy}. 



Against the significant performance and implementation overhead of the CKKS scheme, researchers have proposed a number of optimizations, which are introduced as follows: 

\begin{itemize}
    \item \textbf{Residue Number System (RNS).} In CKKS schemes, the coefficient modulus is extremely large, which can be over 1000 bits. Therefore, the CKKS scheme based on RNS has been proposed\cite{cheon2019fullrns}. This CKKS scheme decomposes a polynomial into multiple smaller ciphertexts with smaller coefficient moduli.
    \label{sec:background:ckks_ksw}
    \item \textbf{Hybrid KeySwitch. } KeySwitch is used in HMult and HRotate to ensure that the ciphertext remains decryptable using the original secret key. KeySwitch is a time-consuming operation in CKKS. To further reduce the performance overhead of KeySwitch, Hybrid KeySwitch\cite{han2020better} is proposed, which reduces the temporary moduli in KeySwitch. A detailed procedure of Hybrid KeySwitch is shown in Algorithm \ref{alg:background:ckks_keyswitch}. 
\end{itemize}

\begin{algorithm}[t]
	\caption{\texttt{TFHE PBS} ($c$, $tv$, $bsk$, $ksk$)} 
	\label{alg:background:tfhe_pbs} 
	\begin{algorithmic} [1]
    \REQUIRE{$c$: a LWE ciphertext, $c$ = ($\textbf{a}$, $b$); $tv$: the test \\vector; $bsk$: the bootstrapping key; $ksk$: the key switching key}
        \STATE $\tilde{c}$ = ($\tilde{\textbf{a}}$, $\tilde{b}$) = ModSwitch($c$) // \texttt{ModSwitch}\\
        \STATE ACC$_{0}$ = Rotate($tv$, $b$)\\
        \STATE// \texttt{Blind Rotation} \\
        \FOR{i = 1 to ${n_{lwe}}$}
        \STATE tmp = (Rotate(ACC$_{i-1}$, $\tilde{a}_i$) - ACC$_{i-1}$)
        \STATE tmp = Decompose(tmp, $l_b$)\\
            \STATE// \texttt{External Product}
            \FOR{j = 1 to $(k+1)l_b$}
            \STATE tmp\_acc = tmp\_acc + NTT(tmp[j]) * $bsk$[i][j] \\
            \ENDFOR
        \STATE ACC$_{i}$ = ACC$_{i-1}$ + INTT(tmp\_acc)\\
        \ENDFOR
        \STATE// \texttt{SampleExtract}
        \STATE $c'$ = ($\textbf{a}'$, $b$) = ($a'_{1}$,...,$a'_{kN}$, $b'$) \\~~~~~~~~~~~~~~= SampleExtract(ACC$_{n}$, 0)\\
        \STATE// \texttt{TFHE KeySwitch}
        \STATE $\textbf{a}''$ = Decompose($\textbf{a}'$, $l_k$)
        \STATE $c''$ = (0,...,0,$b'$) - $\sum_{i=0}^{kN} \sum_{j=0}^{l_k}$ ${\textbf{a}}''_{i}$[j] * $ksk$[i][j]\\
        \RETURN $c''$
	\end{algorithmic}  
\end{algorithm}

\subsection{TFHE scheme}


TFHE is a logic FHE, which encrypts one message into a ciphertext and can enable logic operations including comparison. Here, we briefly introduce the ciphertexts, kernels, and operations in TFHE.

\textbf{GLWE and GGSW ciphertexts. }
GLWE ciphertexts are a generalized version of RLWE for encrypting polynomial-type messages. A GLWE ciphertext is structured as \( (A_1(x), \ldots, A_k(x), B(x)) \) within \( R_q^{k+1} \). The vector \( (A_1(x), \ldots, A_k(x)) \) is known as the GLWE mask. GGSW ciphertexts extend GLWE and are described as a matrix of polynomials of size \( (k+1) \times ((k+1)l_b) \), where \( l_b \) represents the decomposition level of the GGSW ciphertext.


\textbf{External Product.}
The External Product, a crucial operation in programmable bootstrap (PBS), involves multiplying a GLWE ciphertext by a GGSW ciphertext. Initially, the GLWE ciphertext is decomposed into \(l_b\) smaller parts. These parts, forming a vector of size \((k+1)l_b\), are multiplied by a bootstrapping key matrix \(bsk\), a GGSW ciphertext of dimensions \((k+1)l_b \times (k+1)\).

\textbf{PBS.}
PBS, central to TFHE, facilitates arbitrary function evaluation during bootstrapping as detailed in Algorithm \ref{alg:background:tfhe_pbs}. PBS includes a series of operations: ModSwitch, Blind Rotation, SampleExtract, and TFHE KeySwitch. ModSwitch adjusts the scale of a ciphertext, computing \(\tilde{c} = (\lfloor \frac{2N}{q} \cdot \mathbf{a} \rceil, \lfloor \frac{2N}{q} \cdot b \rceil)\). Blind Rotation transforms an LWE ciphertext into a noise-free GLWE ciphertext through \(n_{lwe}\) iterations of External Products. TFHE KeySwitch performs the transformation \(c'' = (0,...,0,b') - \sum_{i=0}^{kN} \sum_{j=0}^{l_k} \mathbf{a}'_{j} \cdot ksk[i][j]\), where \(ksk\) is comprised of \(kN \times l_k\) LWE ciphertexts in \(Z_q^{n_{lwe}+1}\), with \(l_k\) representing the decomposition level of \(ksk\).

\RebuttalChange{
\textbf{Substituting FFT with NTT. }
While FFT is commonly used in TFHE to accelerate External Product, it is possible to substitute FFT with NTT by selecting a prime modulus \(p\), which satisfies \(p \equiv 1 \mod 2N\) and is chosen to be the closest prime to \(q\)\cite{joye2022liberatetfhe, ye2022fpgatfhe}. This facilitates the reuse of NTT hardware in TFHE.
}

\begin{algorithm}[t]
	\caption{\texttt{Scheme Conversion from CKKS to TFHE} ($ct$)} 
	\label{alg:background:schemeswitch_ckks_to_tfhe} 
	\begin{algorithmic} [1]
    \REQUIRE{$ct$: a RLWE ciphertext ($a$, $b$)}
        \FOR{i = 0 to $n_{\text{slot}}$ - 1}
        \STATE $ct'_{i+1}$ = SampleExtract($ct$, i)~~$\triangleright$~\textbf{Sample Extraction}
        \ENDFOR
        \STATE $lwect$ $\gets$ \{$ct'_{1}$,...,$ct'_{n_{slot}}$\}
        \RETURN $lwect$
	\end{algorithmic}  
\end{algorithm}

\begin{algorithm}[t]
	\caption{PackLWEs($\textbf{ct}$)} 
	\label{alg:background:packlwe} 
	\begin{algorithmic} [1]
    \REQUIRE{$\textbf{ct}$: a vector of RLWE ciphertexts, $\textbf{ct}$ = \{${ct}_{1}$,...,${ct}_{n_{slot}}$\}}
        \IF{$n_{slot}$ = 1}
        \STATE $ct$ = $ct_{1}$\\
        \ELSE
        \STATE $ct_{even}$ = PackLWEs(\{$ct_{2j}$\}$_{j \in [1, n_{slot}/2]}$)\\
        \STATE $ct_{odd}$ = PackLWEs(\{$ct_{2j-1}$\}$_{j \in [1, n_{slot}/2]}$)\\
        \STATE $ct$ = ($ct_{even}$ + Rotate($ct_{even}$, $N/n_{slot}$)) + HRotate($ct_{even}$ - Rotate($ct_{odd}$, $N/n_{slot}$), $n_{slot}$ + 1)\\
        \ENDIF
        \RETURN $ct$
	\end{algorithmic}  
\end{algorithm}

\begin{algorithm}[t]
	\caption{\texttt{Scheme Conversion from TFHE to CKKS} ($lwect$)} 
	\label{alg:background:schemeswitch_tfhe_to_ckks} 
	\begin{algorithmic} [1]
    \REQUIRE{$lwect$: $n_{\text{slot}}$ LWE ciphertexts ($\textbf{a}_j$, $b_j$)}
        \STATE Set $ct_j$ = ($a_j$, $b_j$) $\in$ $R^2_q$ for each $j$ $\in$ $[n_{slot}]$, where \\$a_j$ = $\sum_{i \in [N]} \textbf{a}_{j}[i] \cdot X^i$. ~~~~~~~~~~~$\triangleright$~\textbf{Ring Embedding}\\
        \STATE $ct$ = PackLWEs($ct_1$,...,$ct_{n_{slot}}$)~~$\triangleright$~\textbf{Ciphertext Packing}\\
        \FOR{k = 1 to $\log{(N/n_{slot})}$} 
        \STATE $ct$ = $ct$ + HRotate($ct$, $2^{\log{N} - k + 1}$)~~~$\triangleright$~\textbf{Field Trace}
        \ENDFOR
        \RETURN $ct$
	\end{algorithmic}  
\end{algorithm}


\subsection{The Scheme Conversion Between CKKS and TFHE}
\label{sec:background:scheme-conversion}
Scheme Conversion is used to convert between arithmetic FHE and logic FHE. Taking the conversion between CKKS and TFHE as an example, we briefly introduce the Scheme Conversion algorithm.

\textbf{LWE and RLWE ciphertext. }
LWE ciphertext is used as the basic ciphertext in TFHE, while RLWE is used in CKKS. LWE ciphertext encrypts a scalar message, while RLWE ciphertext packs multiple plaintexts. 





\textbf{Scheme Conversion from CKKS to TFHE\cite{chen2021homoconversion}.}
Scheme Conversion from CKKS to TFHE converts an RLWE ciphertext into multiple LWE ciphertexts, as shown in Algorithm \ref{alg:background:schemeswitch_ckks_to_tfhe}. The procedure includes \(n_{\text{slot}}\) multiple SampleExtract operations, where each operation extracts a specific coefficient from the message polynomial encrypted in the RLWE ciphertext.


\textbf{Scheme Conversion from TFHE to CKKS\cite{chen2021homoconversion}.}
Scheme Conversion from TFHE to CKKS packs multiple LWE ciphertexts into a single RLWE ciphertext, as shown in Algorithm \ref{alg:background:packlwe} and \ref{alg:background:schemeswitch_tfhe_to_ckks}. The transformation involves three key steps: Ring Embedding, Ciphertext Packing, and Field Trace. Ring Embedding first transforms an LWE ciphertext into an RLWE format. Ciphertext Packing then merges multiple RLWE ciphertexts using Rotate and HRotate techniques. Finally, Field Trace modifies the combined ciphertext by eliminating unused coefficients of the plaintext polynomial, facilitating further homomorphic operations. Scheme Conversion TFHE to CKKS includes operations such as Rotate and HRotate. Introduced respectively in the TFHE and CKKS schemes, these operations enable the integration and reuse of methodologies from both schemes.


\RebuttalChange{Note that, although there are already several FHE conversion schemes\cite{lu2021pegasus,al2022openfhe,chen2021homoconversion}, we take the scheme from Ref \cite{chen2021homoconversion} as the example to fulfill scheme conversion in this paper, since this algorithm achieves a higher precision than other schemes. Nonetheless, since the Scheme Conversion algorithms (such as OpenFHE\cite{al2022openfhe} and Pegasus\cite{lu2021pegasus}) can also be decomposed into CKKS and TFHE operations, {\SolutionName} can still support these algorithms.}

\section{Motivation}



\RebuttalChange{This section first introduces the requirements for various FHE schemes. Subsequently, we introduce the challenges for using a single type of hardware to support the CKKS, TFHE and the Scheme Conversion between them.} Then, we introduce opportunities for optimization.

\subsection{\RebuttalChange{The necessities for supporting multi FHE schemes}}


\RebuttalChange{Different applications own various characteristics, such as data-intensive, logic-intensive, and a mix of both. Therefore, different kinds of computation are required to meet the various requirements. For example, linear computation is considered as the suitable choice for data-intensive applications due to its SIMD support, while it is less efficient for logic-intensive applications. When considering the FHE schemes, the schemes with linear operation support (i.e., BGV, BFV, CKKS) are proper choice for data-intensive applications, while the FHE scheme with logical operation support (i.e., TFHE) is suitable for logic-intensive applications. Moreover, for the FHE applications with mixed requirements (i.e., FHE database), both linear FHE scheme and logical FHE scheme are simultaneously required. Therefore, for a real-world system, it is important to meet the requirements of different kinds of applications and thereby require efficient support for multi FHE schemes.
}

\subsection{\textbf{Challenge 1:} Various polynomial lengths in CKKS and TFHE}



\begin{figure}
    \centering
    \includegraphics[width=\linewidth]{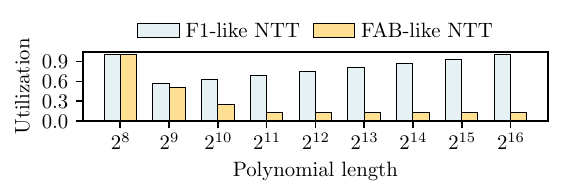}
    \caption{Utilization of F1-like NTT and FAB-like NTT when computing NTT of varying lengths. {For a fair analysis, both the F1-like NTT and the FAB-like NTT are configured with comparable modular multipliers. The F1-like NTT includes eight stages of butterfly units and processes 256 elements in parallel per cycle. In contrast, the FAB-like NTT consists of a single butterfly stage capable of processing 2048 elements in parallel per cycle. \RebuttalChange{Both NTT employ radix-2 NTT and support four-step NTT. } The utilization rate is computed considering a single butterfly stage as the finest granularity.}}
    \label{fig:motiv:ntt_util}
\end{figure}

In TFHE, due to security considerations and a relatively small coefficient modulus, a polynomial length ranging from 256 to 4096 is sufficient. In contrast, in CKKS, particularly for configurations that include Bootstrapping, \(N\) is usually set to $2^{16}$. We investigate the utilization of F1-like NTT and FAB-like NTT under various polynomial lengths in Figure \ref{fig:motiv:ntt_util}. The result demonstrates that a fixed hardware design cannot achieve high utilization across various polynomial lengths: 1) F1-like NTT achieves its highest utilization when $N$ = $2^{16}$, while the utilization starts to decrease when $N$ decreases; 2) FAB-like achieve its highest utilization when $N$ = $2^8$, while the utilization starts to decrease when $N$ increases. Therefore, it would be hard for specific hardware to efficiently support the NTT across various polynomial lengths, when supporting CKKS and TFHE schemes. This will incur low utilization and thereby performance degradation.


\label{sec:motiv:ntt_design}

\subsection{\textbf{Challenge 2:} Imbalanced computation workload}

\begin{figure}[t]
    \vspace{-10pt}
    \centering
    \includegraphics[width=\linewidth]{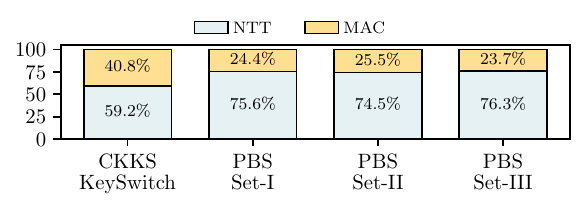}
    \caption{The computational amount breakdown of NTT and MAC operation in CKKS KeySwitch ($L$ = 23, dnum = 3) and TFHE PBS. }
    \vspace{-10pt}
    \label{fig:motiv:ntt-and-mac-breakdown}
\end{figure}


As introduced in Section \ref{sec:background:ckks_ksw}, KeySwitch is a pivotal operation in the CKKS scheme, employed in both HRotate and HMult, whereas PBS is a fundamental operation for arbitrary function evaluation in TFHE. KeySwitch consists primarily of NTT, BConv, and Inner Product computations, where NTT and BConv can be computed using MAC. 
PBS predominantly involves external products that include NTT and MAC computations. 
In the FHE accelerator, the implementation of the NTT unit commonly utilizes multiple butterfly stages, while the computation of MAC utilizes a systolic array. Due to this difference, there are dedicated computational components for NTT and MAC, respectively. However, as shown in Figure \ref{fig:motiv:ntt-and-mac-breakdown}, due to the computation of FHE scheme, there is workload imbalance: 1) when running CKKS KeySwitch, the computational load of NTT accounts for 59.2\% of the total in CKKS KeySwitch, while MAC constitutes 40.8\%; 2) For PBS, NTT represents on average 75.5\% of the total computational load for PBS, with MAC comprising the remaining 24.5\%. In this situation, the computation of NTT becomes the computational bottleneck, where the NTT unit is running with full workloads while the systolic array remains idle. The above analysis suggests that it is difficult to use the same hardware configuration to efficiently support CKKS and TFHE. 

\subsection{Opportunities}



For the analysis and challenges presented, we conducted a thorough investigation and arrived at the following observations and opportunities:
\begin{itemize}
    \item We observe that it is possible to substitute FFT with NTT in TFHE by selecting a proper value for $q$\cite{joye2022liberatetfhe, ye2022fpgatfhe}. Given that prior CKKS accelerators \cite{kim2023sharp, kim2022ark, samardzic2022craterlake} implement NTT units, this provides the opportunity to reuse the NTT unit of CKKS accelerators in the computation of TFHE. 
    \item By comparing F1-like NTT and FAB-like NTT, we observe that the optimal polynomial lengths supported by NTT units vary according to the number of butterfly stages. This inspires us to deploy heterogeneous components to support NTT with various polynomial lengths.
    \item As discussed in Section \ref{sec:background:ckks_kernels}, both BConv and IP can be executed using a systolic array. Furthermore, the computation pattern for NTT also aligns similarly with that of a systolic array\cite{nejatollahi2020santt}. This alignment provides an opportunity to design a configurable component capable of supporting both NTT and MAC computations. By using parts of this component for NTT computations, we can ensure a balanced computational workload across both CKKS and TFHE schemes, thereby enhancing performance.
\end{itemize}

In this work, we propose a novel and configurable accelerator named {\SolutionName}, which is designed to provide efficient and flexible support for the computation of the CKKS, TFHE scheme, and the conversion between them. 

\section{Design}

\begin{figure}[t]
    \centering
    \includegraphics[width=\linewidth]{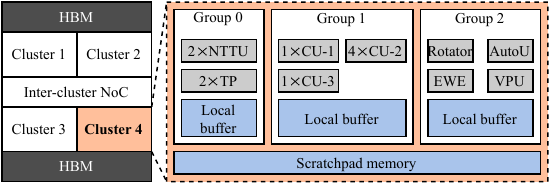}
    \caption{Overall Architecture of \emph{\SolutionName}. NTTU denotes the NTT unit. TP denotes transpose unit. CU-$x$ denotes a configurable unit with $x$-column PEs. }
    \label{fig:design:overall-arch}
\end{figure}

\subsection{Overview}

Figure \ref{fig:design:overall-arch} presents the overall architecture of \emph{\SolutionName}, which mainly contains the following components:

\begin{itemize}

    \item \textbf{Cluster}. The cluster functions as a high-performance and configurable computing engine, efficiently supporting the operations of the CKKS and TFHE schemes and facilitating conversions between them. As illustrated in the right section of Figure \ref{fig:design:overall-arch}, each cluster has a scratchpad memory and three heterogeneous groups: Group 0, Group 1, and Group 2. Group 0 comprises two transpose units (TP) and two NTT units (denoted as NTTU). 
    Group 1 includes a set of configurable units. These configurable units contain multiple columns of processing elements (PE). Here, we denote the configurable unit as CU-$x$, which means a configurable unit with $x$-column PEs. Besides, we use CU to refer to all kinds of CU-$x$. Group 1 includes six CUs, designated one CU-1, four CU-2, and one CU-3. 
    CU is pivotal in \emph{\SolutionName}, supporting both NTT and MAC computations (detailed in Section \ref{sec:design:cu}). 
    Group 2 contains a Rotator, an automorphism unit (AutoU), an element-wise engine (EWE), and a vector processing unit (VPU). 

    \item \textbf{Inter-cluster NoC}. This network-on-chip (NoC) enables all-to-all data exchange among different clusters, playing a crucial role in switching between two data layouts (detailed in Section \ref{sec:design:data_layout}). \RebuttalChange{This NoC employs a fully-connected topology. }

    \item \textbf{On-chip memory}. There are two types of on-chip memory, namely scratchpad memory and local buffer. Scratchpad memory is shared across the groups in a cluster, while the local buffer is shared across the functional units in a group. Each group has a local buffer, while each has a scratchpad memory. 

    \item \textbf{HBM}. \emph{\SolutionName} facilitates two HBM2 interfaces for off-chip data exchange, providing a total bandwidth of 1TB/s.

\end{itemize}

\subsection{NTTU structure}

\begin{figure}
    \centering
    \includegraphics[width=\linewidth]{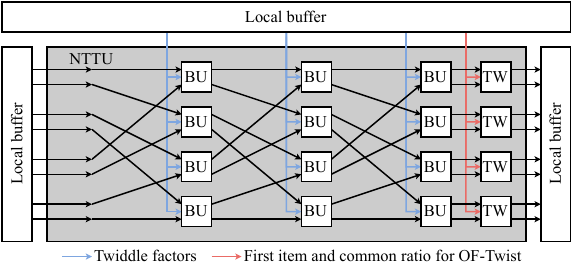}
    \caption{\RebuttalChange{The structure of NTTU. We denote the number of rows of the BU array as $M$. In the default configuration of \emph{\SolutionName}, we set $M$ as 128, and NTTU processes 256 elements each cycle. For simplicity, here we take $M$ = 4 as an example. }}
    \label{fig:design:bfu}
\end{figure}


\RebuttalChange{As shown in Figure \ref{fig:design:bfu}, The NTTU supports NTT computations, which contains multiple butterfly units (BUs) with $\log_2{(2M)}$ stages and one-stage twisting units (TWs). BU computes the butterfly operation, while TW computes the twisting operation in four-step NTT. In the design of the BU array in NTTU, We employ constant-geometry\cite{chen2014high} NTT\cite{pease1968adaptation}, which can maintain a consistent access pattern for the computation of BUs in each stage. 
\RebuttalChange{To minimize memory bandwidth requirements, we implement on-the-fly twisting factor generation (OF-Twist), similar to the approach in ARK \cite{kim2022ark}. }
NTTU enables the computation of $2M$-point NTT and works in a fully pipelined manner, processing $2M$ elements per cycle. 
}

\RebuttalChange{For the transmission of auxiliary data, the twiddle factors are fed in BUs from the local buffer as shown in Figure \ref{fig:design:bfu}. For twisting factors, only the first item and common ratios are fed in TW for the on-the-fly twisting factor generation. }

\RebuttalChange{
In \emph{\SolutionName}, we set $M$ as 128, which means an NTTU can facilitate the computation of 256-point NTT. By collaborating with CU, NTTU can compute the NTT of polynomial lengths ranging from 256 to 65536. 
}

\subsection{Configurable Unit}
\label{sec:design:cu}

\subsubsection{Architecture of Configurable Unit}

\begin{figure}[t]
    \centering
    \includegraphics[width=\linewidth]{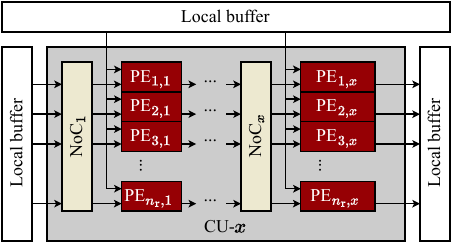}
    \caption{CU-$x$ architecture}
    \label{fig:design:cu-arch}
\end{figure}

To enhance computational efficiency across varying polynomial sizes and TFHE parameter sets, we propose a configurable unit named CU. 
Figure \ref{fig:design:cu-arch} presents the structure of CU-$x$, which consists of $x$ columns and $n_r$ rows of PE. These PE are interconnected via a Network-on-Chip (NoC) that facilitates the specific access patterns required for NTT computations and systolic arrays. \RebuttalChange{The NoC can be configured as a 2D-mesh topology for systolic array and butterfly topology for NTT. } \RebuttalChange{Similar to NTTU, our implementation adopts the butterfly topology based on the constant-geometry algorithm, maintaining consistent access patterns across all butterfly stages.} This approach simplifies the NoC design and enhances the configurability of the CU. By employing this approach, the area overhead of the NoC constitutes only 0.2\% of the total area of CU-$x$.

\subsubsection{PE structure}

\begin{figure}[t]
    \centering
    \includegraphics[width=\linewidth]{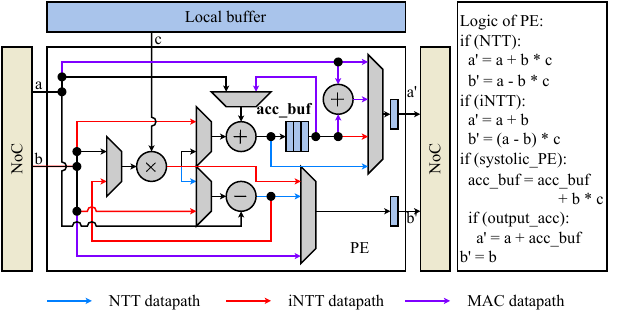}
    \caption{PE structure}
    \label{fig:design:pe-structure}
\end{figure}

\begin{figure*}[t]
    \centering
    \includegraphics[width=\linewidth]{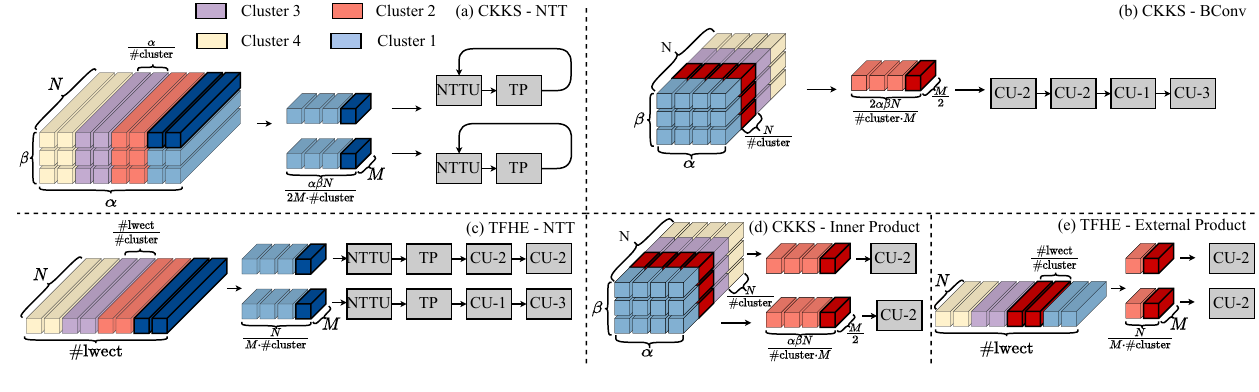}
    \caption{Mapping strategy and data layout for kernels in different schemes, including a) NTT in CKKS; b) BConv in CKKS; c) NTT in TFHE; d) Inner Product in CKKS; e) External Product in TFHE. }
    \label{fig:design:data-map}
\end{figure*}

While the NoC in CU-$x$ provides different access patterns for NTT and systolic array, the PEs in CU-$x$ provide different compute patterns for NTT and MAC operations. Figure \ref{fig:design:pe-structure} presents the structure and datapath of PE. This PE supports multiple computing patterns. As shown in Figure \ref{fig:design:pe-structure}, the arrow with different colors denotes a kind of datapath for a specific computation, while the arrow with black color denotes the shared datapath for more than two computations. The PE can be configured to compute NTT, iNTT, and MAC. 

Each CU works in a fully-pipelined manner. With the capability of PE, CU supports both NTT and MAC computations. Note that the throughput is different when CU is computing NTT and when CU is computing MAC. When computing NTT, each CU processes $2n_r$ elements per cycle. When computing MAC, each CU processes $n_r$ elements per cycle. In \emph{\SolutionName}, $n_r$ is set to 128, and each CU processes 256 elements per cycle when computing NTT and processes 256 elements per cycle when computing MAC.



\subsection{Transpose unit, element-wise engine, automorphism unit, Rotator, vector processing unit}


\textbf{Transpose unit (TP). }
The TP manages the transposition of polynomials ranging in size from 512 to 65536. TP follows the same design of \cite{samardzic2021f1}, containing multiple stages of quad-swap units. Unused stages are bypassed for smaller polynomials. 

\textbf{Element-wise engine (EWE). }
The design of EWE follows \cite{kim2023sharp},  which can handle modular operations such as ModAdd and ModMult.

\textbf{Automorphism unit (AutoU). }
The design of AutoU follows \cite{kim2022ark}, which includes multiple stages of shuffle units for automorphism computations. 

\textbf{Rotator. }
Rotator enables vector rotation and SampleExtract operations, featuring buffers, a vector rotate unit, and a vector negation unit. 

\textbf{Vector processing unit (VPU). }
VPU handles Modulus Switch and KeySwitch operations, following the design of Morphling\cite{prasetiyo2024morphling}.  

Among all the above components, each EWE can process 512 elements in parallel per cycle, whereas other components handle 256 elements per cycle.

\subsection{\RebuttalChange{Efficient support for NTT with various polynomial lengths}}
\RebuttalChange{When computing $2M$-point NTT, the TWs are bypassed. When computing the NTT with a larger polynomial length than $2M$, we employ the four-step NTT method\cite{bailey1990ffts}, which divides NTT into two smaller NTTs, namely phase-1 NTT and phase-2 NTT. Besides, we develop two computing strategies for different cases. When computing $4M^2$-point NTT, NTTU computes both the phase-1 NTT and phase-2 NTT. When computing NTT with the polynomial length ranging from $4M$ to $2M^2$, the phase-1 NTT is computed by NTTU, while the phase-2 NTT is computed by CU (detailed in Section \ref{sec:design:cu}). }

\subsection{\RebuttalChange{Efficient support for imbalanced FHE workloads}}
\RebuttalChange{In this section, we introduce the data mapping of kernels for different schemes on \emph{\SolutionName}, focusing on NTT, BConv, Inner Product, and External Product---key operations in both CKKS and TFHE.}
\RebuttalChange{
Figure \ref{fig:design:data-map} presents the data mapping for kernels in CKKS (N = 65536) and TFHE (N = 4096). Our strategy prioritizes fulfilling NTT requirements first. Subsequently, unutilized CUs are allocated for the computations of BConv, Inner Product, and External Product. Specifically, in the CKKS scheme, two NTTUs and two TPs are dedicated to NTT computation, as shown in Figure \ref{fig:design:data-map}(a). For BConv computations, one CU-1, one CU-3, and one CU-4 are utilized, as shown in Figure \ref{fig:design:data-map}(c). Notably, we modify the computation of the Inner Product by deploying two CU-2s instead of using the EWE, alleviating the computational load on the EWE and enhancing CKKS performance. 
}

\RebuttalChange{
For TFHE, as shown in Figure \ref{fig:design:data-map}(c), NTTU, CU-1, CU-3, and two CU-2 are deployed for NTT. This deployment allows for parallel processing of two NTTs, achieving high throughput and high utilization. Besides, the External Product is computed using two CU-2s, as shown in Figure \ref{fig:design:data-map}(d).
}

\RebuttalChange{
Through this approach, \emph{\SolutionName} leverages the configurability of CU-$x$ to dynamically meet the diverse requirements of NTT and MAC operations across varying polynomial sizes and different schemes. This method significantly enhances performance across various scenarios while maintaining a modest area.}

\subsection{\RebuttalChange{Support for scheme conversion}}

\RebuttalChange{As introduced in Section \ref{sec:background:scheme-conversion}, the computation of scheme conversion reuses operations from CKKS and TFHE. The scheme conversion from CKKS to TFHE purely contains SampleExtract, which is performed by Rotator. The scheme conversion from TFHE to CKKS mainly includes HRotate, and Rotate. Rotate is performed by Rotator. HRotate is performed by AutoU, NTTU, CU, and EWE. }

\subsection{\RebuttalChange{Hardware control}}

\RebuttalChange{For the components except CU, these components are all separate and only support a single kernel from CKKS or TFHE. For CUs, despite being utilized for the computation of different schemes, Trinity does not simultaneously execute the operations from different schemes. Therefore, CUs do not cause contention and do not induce additional control overheads.}

\subsection{\RebuttalChange{Data layout}}


\label{sec:design:data_layout}
\RebuttalChange{In terms of data layout, \emph{\SolutionName} employs two data layout patterns: limb-wise and slot-wise, aligning with the configurations in ARK \cite{kim2022ark}. As depicted in Figure \ref{fig:design:data-map}, limb-wise layout is adopted for NTT computations in CKKS and all operations in TFHE, whereas slot-wise layout is used for BConv and Inner Product computations in CKKS. Switching between these layouts is managed via the inter-cluster NoC.}

\subsection{On-chip memory system and on-chip network}

The on-chip memory system of \emph{\SolutionName} comprises multi-level memory, which includes scratchpads and local buffers. The scratchpad facilitates data exchange with the HBM, other clusters, and within the cluster groups, coordinating data transfers to other clusters via the inter-cluster NoC and to different groups through the local buffers. Each local buffer, equipped with 256 lanes, features vectorized memory with each lane comprising five single-ported banks of 36-bit width. Each bank can store two polynomials of length 65536, providing a total capacity of 2.81 MB and a total bandwidth of 11.25 TB/s per local buffer. Similarly, the scratchpad comprises 256 lanes containing four single-ported banks, each 36-bit wide. This configuration allows each bank to store 40960 items, adequately supporting storage requirements for polynomials, evk, bsk, and ksk, with a total capacity of 45 MB and a bandwidth of 9 TB/s.

The on-chip network in \emph{\SolutionName} can be categorized into three types. The first type is the inter-cluster NoC, which facilitates all-to-all data exchange among different clusters and switching between limb-wise and slot-wise data layouts. 
The second type is the inter-group NoC, which facilitates data exchange among the groups in a cluster. 
The third type is the intra-group NoC, which facilitates data exchange between different computational units within a group.
The fourth type is the NoC inside the CU, which facilitates data exchange between consecutive butterfly stages.

\begin{figure}[t]
    \centering
    \includegraphics[width=\linewidth]{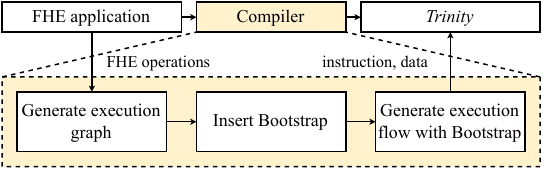}
    \caption{Workload allocation procedure in \emph{\SolutionName}. }
    \label{fig:design:workload-alloc}
\end{figure}
\subsection{Workload Allocation}


\RebuttalChange{Figure \ref{fig:design:workload-alloc} illustrates the procedure of workload allocation for \emph{\SolutionName}. For one FHE workload based on CKKS, TFHE, or Hybrid Schemes, it will be firstly decomposed as the kernel flow. Then, the kernel flow is carefully scheduled to eliminate the hardware hazards and guarantee hardware utilization. After that, the accelerator executes the scheduled kernel flow without distinguishing which FHE scheme the kernel comes from. In this way, Trinity enables the support for the FHE applications based on both single schemes and hybrid schemes, and even supports for simultaneous execution of multiple FHE applications, without hardware switching overhead.}


\section{Methodology}

\begin{table}[t]
    \caption{The configuration of \emph{\SolutionName}. }
    \centering
    \begin{tabular}{l|l}
    \toprule[1pt]
    \textbf{Parameter} & \textbf{Value} \\
    \midrule[0.5pt]
    word size & 36bit\\
    \# of cluster      & 4  \\
    \makecell[lt]{Capacity of \\scratchpad memory}     &  180MB \\
    $n_r$ of CU     &  128 \\
    $M$ of NTTU     &  128 \\
    \#$_{\text{butterfly stage}}$ of NTTU     &  8 \\
    \# of HBM2 stacks & 2 \\
    \bottomrule[1pt]
    \end{tabular}
    \label{tab:method:hard_config}
\end{table}

\begin{table}[t]
    \caption{\RebuttalChange{CKKS and TFHE parameter sets.} }
    \centering
    \begin{tabular}{lrrrrr}
    \toprule[1pt]
    \multicolumn{6}{c}{\textbf{CKKS}} \\
         & N & $L$ & dnum & & $\lambda$ \\
    \midrule[0.5pt]
    \RebuttalChange{default}     & 65536 & 35 & 3 & & 128-bit \\
    \midrule[1pt]
    \multicolumn{6}{c}{\textbf{TFHE}} \\
         & N & n$_{\text{lwe}}$ & k & $l_b$ & $\lambda$ \\
    \midrule[0.5pt]
    Set-I     & 1024 & 500 & 1 & 2 & 80-bit  \\
    Set-II     & 1024 & 630 & 1 & 3 & 110-bit  \\
    Set-III     & 2048 & 592 & 1 & 3 & 128-bit \\
    \bottomrule[1pt]
    \end{tabular}
    \label{tab:method:ckks_tfhe_param_set}
\end{table}

\begin{table}[t]
\centering
\caption{\RebuttalChange{Comparison Scheme}}
\label{tab:method:comparison}
\begin{tabular}{lll}
\toprule[1pt]
\multicolumn{1}{l}{\textbf{Type}} & \multicolumn{1}{l}{\textbf{Design}}                   & \textbf{Description} \\ 
\midrule[0.5pt]
\multicolumn{3}{c}{\textbf{CKKS}}                                                                                 \\ 
\midrule[0.5pt]
\multicolumn{1}{l|}{CPU}  & \multicolumn{1}{l|}{Baseline-CKKS\cite{samardzic2022craterlake}}                     &    AMD Ryzen 3975WX                  \\ 
\multicolumn{1}{l|}{GPU}  & \multicolumn{1}{l|}{TensorFHE\cite{fan2023tensorfhe}}                         &   NTT with TCUs                   \\ 
\multicolumn{1}{l|}{ASIC} & \multicolumn{1}{l|}{CraterLake\cite{samardzic2022craterlake}}                        &  \makecell[tl]{1$\times$CRB, 2$\times$NTT, 1$\times$Auto, \\5$\times$Mul, 5$\times$Add  }                 \\ 
\multicolumn{1}{l|}{ASIC} & \multicolumn{1}{l|}{BTS\cite{kim2022bts}}                               &   \makecell[tl]{2048$\times$PE, each with\\ 1 ModMult, 1 MMAU, 1NTTU}                   \\ 
\multicolumn{1}{l|}{ASIC} & \multicolumn{1}{l|}{ARK\cite{kim2022ark}}                               &   \makecell[tl]{4 clusters each with 1 NTTU, \\1 BConvU, 1 AutoU, 2 MADU}                   \\ 
\multicolumn{1}{l|}{ASIC} & \multicolumn{1}{l|}{SHARP\cite{kim2023sharp}}                             &   \makecell[tl]{4 clusters, each with 1 NTTU,\\ 1 BConvU, 1 AutoU, 1 EWE}                   \\ 
\multicolumn{1}{l|}{ASIC} & \multicolumn{1}{l|}{\emph{\SolutionName}} &     \makecell[tl]{4 clusters, each with 2 NTTU,\\ 1 CU-1, 4 CU-2, 1 CU-3, \\1 AutoU, 1 Rotator, 1 VPU,\\ 1 EWE}                 \\ 
\midrule[0.5pt]
\multicolumn{3}{c}{\textbf{TFHE}}                                                                                 \\ 
\midrule[0.5pt]
\multicolumn{1}{l|}{CPU}  & \multicolumn{1}{l|}{Baseline-TFHE\cite{prasetiyo2024morphling}}                     &   Intel Xeon
Platinum                   \\ 
\multicolumn{1}{l|}{GPU}  & \multicolumn{1}{l|}{NuFHE\cite{nucypher2020nufhe}}                                  &     Nvidia Titan RTX                 \\ 
\multicolumn{1}{l|}{ASIC} & \multicolumn{1}{l|}{Matcha\cite{jiang2022matcha}}                            &  \makecell[tl]{32$\times$IFFT, 8$\times$FFT, \\160$\times$Mult, 192$\times$Add}                    \\ 
\multicolumn{1}{l|}{ASIC} & \multicolumn{1}{l|}{Strix\cite{putra2023strix}}                         &  \makecell[tl]{8$\times$HSC, each with 2 VMA, \\1 IFFT, 1 FFT, 2 Decomp, \\2 Accum, 1 Rotator }                   \\ 
\multicolumn{1}{l|}{ASIC} & \multicolumn{1}{l|}{Morphling\cite{prasetiyo2024morphling}}                         &  \makecell[tl]{8$\times$FFT, 16$\times$ IFFT, \\64$\times$VPE,1$\times$VPU}                    \\ 
\multicolumn{1}{l|}{ASIC} & \multicolumn{1}{l|}{\emph{\SolutionName}} &     -                \\ 
\midrule[0.5pt]
\multicolumn{3}{c}{\textbf{\RebuttalChange{Hybrid scheme}}}                                                                            \\ 
\midrule[0.5pt]
\multicolumn{1}{l|}{CPU}  & \multicolumn{1}{l|}{Baseline-SC\cite{chen2021homoconversion}}                    &       I7-4770K              \\ 
\multicolumn{1}{l|}{{CPU}}  & \multicolumn{1}{l|}{\RebuttalChange{Baseline-Hybrid\cite{bian2023he3db}}}                    &       {Intel Xeon Platinum}             \\ 
\multicolumn{1}{l|}{{System}} & \multicolumn{1}{l|}{\RebuttalChange{SHARP+Morphling}}      &     \makecell[tl]{A system including SHARP \\and Morphling, assuming a 128GB/s \\PCIE 5 connection between them}                \\
\multicolumn{1}{l|}{ASIC} & \multicolumn{1}{l|}{\emph{\SolutionName}}      &     -                \\
\bottomrule[1pt]
\end{tabular}
\end{table}

\subsection{Hardware modeling}

For the evaluation of power, area, and frequency, all the hardware modules in \emph{\SolutionName} are implemented using Verilog and synthesized with the TSMC 7nm Process Design Kit (PDK). 
All the modules are fully pipelined, enabling operation at a frequency of 1GHz. 
For the estimation of SRAM components, we utilize the SRAM components provided by the PDK, all of which are double-pumped. 
For the estimation of NoC, we utilize ORION \cite{kahng2015orion3}, a tool for estimating the area and power of NoC. 
Furthermore, we develop a cycle-accurate simulator to evaluate the performance under full FHE workloads.
Table \ref{tab:method:hard_config} presents the hardware configurations of \emph{\SolutionName} that are used in the experiment. 

\subsection{Benchmark suite}
In this section, we introduce the workloads that are used to compare with other ASIC accelerators. We have categorized the benchmarks we used into three categories: CKKS Benchmark, TFHE Benchmark, and Scheme Conversion Benchmark.
\subsubsection{CKKS Benchmark}
We utilize the following CKKS applications to evaluate the CKKS effectiveness of \emph{\SolutionName}:
\begin{itemize}
    \item \textbf{Packed Bootstrapping\cite{mouchet2020lattigo}}: This example involves the operation of a fully packed bootstrapping. The level consumption of bootstrapping is 15. 
    \item \textbf{Logistics Regression\cite{han2019helr}}: This benchmark involves the training of a binary classification model using logistic regression. The batch size is set to 1024 and the number of training iterations is 32. 
    \item \textbf{ResNet-20\cite{lee2022resnet20}}: This benchmark involves a CNN inference with CIFAR-10 dataset using the CKKS-based ResNet-20 model. The size of the input image is 32 $\times$ 32 $\times$ 3.
\end{itemize}
Note that all the CKKS benchmarks are evaluated using the default CKKS parameter sets in Table \ref{tab:method:ckks_tfhe_param_set}.

\subsubsection{TFHE Benchmark}
For the TFHE effectiveness evaluation, the following applications are utilized as the benchmark:
\begin{itemize}
    \item \textbf{Programmable Bootstrapping (PBS)\cite{chillotti2021pbs}}: This benchmark involves programmable bootstrapping. We evaluate the performance of PBS under Set-I, Set-II, and Set-III, respectively (see in Table \ref{tab:method:ckks_tfhe_param_set}).
    \item \textbf{NN-$x$\cite{chillotti2021pbs}}: This benchmark involves a CNN inference on the MNIST dataset. The size of input images is 28 $\times$ 28. The $x$ denotes the depth of the neural networks. We tested the latency of NN-20, NN-50, and NN-100. The NN on baseline-TFHE is evaluated using 12 threads of the Intel Xeon Platinum 8280 CPU. 
\end{itemize}

\subsubsection{Scheme Conversion Benchmark}
We use the conversion scheme introduced in Ref \cite{chen2021homoconversion} to evaluate \emph{\SolutionName}. This scheme involves the repacking procedure that converts a set of LWE ciphertexts into an RLWE ciphertext. The parameter $n_{\text{slot}}$ denotes the number of LWE ciphertext. Note that, we do not test the conversion from CKKS to TFHE operation, since this operation is simple, and only contains multiple \emph{SampleExtract} operations. For ease of comparison, we set $N$ = $2^{14}$ and $L$ = 8 in this benchmark, which is consistent with Ref \cite{chen2021homoconversion}.

\RebuttalChange{
\subsubsection{Hybrid scheme Benchmark}
\textbf{HE3DB-$x$\cite{bian2023he3db}}: This benchmark homomorphically performs Query 6 in the TPC-H benchmark\cite{noauthor_tpc_2022}, which is a kind of typical queries. This benchmark involves filter and aggregation, which are respectively executed in TFHE and CKKS domains. Scheme conversion is performed between filter and aggregation. We denote this benchmark as HE3DB-$x$, where $x$ denotes the number of entries to be queried. We evaluate the latency of HE3DB-4096 and HE3DB-16384. The HE3DB on baseline-Hybrid is evaluated using a single thread of Intel Xeon Platinum 8280. To compare with SOTA accelerators, we assume a system named SHARP+Morphling, where its configuration is listed in Table \ref{tab:method:comparison}. In SHARP+Morphling, we assume the TFHE-to-CKKS conversion is performed on SHARP and the CKKS-to-TFHE conversion is performed on Morphling. 
}

Together with the above benchmark, we can make a thorough analysis of the performance and efficiency of \emph{\SolutionName} when executing full-FHE applications based on CKKS, TFHE and Scheme Conversion between them. The designs used for comparison are shown in Table \ref{tab:method:comparison}. 

\begin{table}[t]
\centering
\caption{\RebuttalChange{Performance for CKKS workloads (ms).}}
\label{tab:result:app-perf}
\begin{tabular}{llll}
\toprule[1pt]
\multicolumn{1}{l}{\textbf{Scheme}}  & \multicolumn{1}{l}{\textbf{Bootstrap}}  &  \multicolumn{1}{l}{\textbf{HELR}}   & \multicolumn{1}{l}{\textbf{ResNet-20}}                      \\ 
\midrule[0.5pt]
\multicolumn{1}{l}{Baseline-CKKS}  & \multicolumn{1}{r}{17.2s}                     &    \multicolumn{1}{r}{356s}   & \multicolumn{1}{r}{23min}                  \\ 
\multicolumn{1}{l}{TensorFHE\cite{fan2023tensorfhe}}  & \multicolumn{1}{r}{421.8}                         &     \multicolumn{1}{r}{220}   & \multicolumn{1}{r}{4,939}                 \\ 
\multicolumn{1}{l}{F1\cite{samardzic2021f1}} & \multicolumn{1}{r}{-}                                &      \multicolumn{1}{r}{639}   & \multicolumn{1}{r}{2,693}                \\ 
\multicolumn{1}{l}{CraterLake\cite{samardzic2022craterlake}} & \multicolumn{1}{r}{3.91}                        &      \multicolumn{1}{r}{119.52}   & \multicolumn{1}{r}{249.45}                \\ 
\multicolumn{1}{l}{BTS\cite{kim2022bts}} & \multicolumn{1}{r}{22.88}                               &    \multicolumn{1}{r}{28.4}   & \multicolumn{1}{r}{1,910}                  \\  
\multicolumn{1}{l}{ARK\cite{kim2022ark}} & \multicolumn{1}{r}{3.52}                               &    \multicolumn{1}{r}{7.42}   & \multicolumn{1}{r}{125}                  \\ 
\multicolumn{1}{l}{SHARP\cite{kim2023sharp}} & \multicolumn{1}{r}{3.12}                             &  \multicolumn{1}{r}{2.53}   & \multicolumn{1}{r}{99}                    \\ 
\multicolumn{1}{l}{\RebuttalChange{\emph{\SolutionName}}} & \multicolumn{1}{r}{\Tofill{1.92}} &      \multicolumn{1}{r}{\Tofill{1.37}}   & \multicolumn{1}{r}{\Tofill{89}}                \\ 
\bottomrule[1pt]
\end{tabular}
\end{table}

\begin{table}[t]
\centering
\caption{Throughput for TFHE PBS (OPS).}
\label{tab:result:tfhe-pbs-perf}
\begin{tabular}{llll}
\toprule[1pt]
\multicolumn{1}{l}{\textbf{Scheme}}  & \multicolumn{1}{r}{\textbf{Set-I}}  &  \multicolumn{1}{r}{\textbf{Set-II}}   & \multicolumn{1}{r}{\textbf{Set-III}}                      \\
\midrule[0.5pt]
\multicolumn{1}{l}{Baseline-TFHE\cite{zama2022Concrete}}  & \multicolumn{1}{r}{63}                     &       \multicolumn{1}{r}{36}   & \multicolumn{1}{r}{12}               \\ 
\multicolumn{1}{l}{GPU}  & \multicolumn{1}{r}{2,500}                                  &              \multicolumn{1}{r}{550}   & \multicolumn{1}{r}{-}        \\ 
\multicolumn{1}{l}{Matcha\cite{jiang2022matcha}} & \multicolumn{1}{r}{10,000}    &                 \multicolumn{1}{r}{-}   & \multicolumn{1}{r}{-}                             \\ 
\multicolumn{1}{l}{Strix\cite{putra2023strix}} & \multicolumn{1}{r}{74,696}                         &       \multicolumn{1}{r}{39,600}   & \multicolumn{1}{r}{21,104}               \\ 
\multicolumn{1}{l}{Morphling\cite{prasetiyo2024morphling}} & \multicolumn{1}{r}{147,615}                         &       \multicolumn{1}{r}{78,692}   & \multicolumn{1}{r}{41,850}               \\ 

\multicolumn{1}{l}{Morphling$_{\text{1GHz}}$} & \multicolumn{1}{r}{\Tofill{123,012}} &      \multicolumn{1}{r}{\Tofill{65,576}}   & \multicolumn{1}{r}{\Tofill{34,875}}   \\
\multicolumn{1}{l}{\emph{\SolutionName}-TFHE$_{\text{w/o CU}}$} & \multicolumn{1}{r}{\Tofill{83,333}} &      \multicolumn{1}{r}{\Tofill{49,603}}   & \multicolumn{1}{r}{\Tofill{26,393}}                \\ 
\multicolumn{1}{l}{\emph{\SolutionName}-TFHE$_{\text{w/ CU}}$} & \multicolumn{1}{r}{\Tofill{150,015}} &      \multicolumn{1}{r}{\Tofill{85,034}}   & \multicolumn{1}{r}{\Tofill{45,246}}                \\ 
\multicolumn{1}{l}{\emph{\SolutionName}} & \multicolumn{1}{r}{\Tofill{600,060}} &      \multicolumn{1}{r}{\Tofill{340,136}}   & \multicolumn{1}{r}{\Tofill{180,987}}                \\ 
\bottomrule[1pt]
\end{tabular}
\end{table}

\subsection{Compared scheme setting}

For a thorough analysis for \emph{\SolutionName}, we established several compared schemes.

\textbf{CKKS. }
For CKKS, we established one compared scheme named \emph{\SolutionName}-CKKS$_{\text{IP-use-EWE}}$. The difference between \emph{\SolutionName}-CKKS$_{\text{IP-use-EWE}}$ and \emph{\SolutionName} is that \emph{\SolutionName}-CKKS$_{\text{IP-use-EWE}}$ uses EWE for the computation of IP, whereas \emph{\SolutionName} uses some of the CU for this purpose. 

\textbf{TFHE. }
For TFHE, we set up three compared schemes: Morphling$_{\text{1GHz}}$, \emph{\SolutionName}-TFHE$_{\text{w/ CU}}$ and \emph{\SolutionName}-TFHE$_{\text{w/o CU}}$. Morphling$_{\text{1GHz}}$ has its frequency set to 1GHz, compared with the original design of Morphling. Both \emph{\SolutionName}-TFHE$_{\text{w/ CU}}$ and \emph{\SolutionName}-TFHE$_{\text{w/o CU}}$ maintain the same level of parallelism in the NTT unit as the FFT unit in Morphling. \emph{\SolutionName}-TFHE$_{\text{w/ CU}}$ retains the architecture of \emph{\SolutionName}, except it scales down the parallelism. \emph{\SolutionName}-TFHE$_{\text{w/o CU}}$ is a fixed design that includes NTT units and a systolic array (SA) but lacks the capability for flexible mapping. The depth of SA in TFHE$_{\text{w/o CU}}$ is 12, which is consistent with the total depth of all CU in \emph{\SolutionName}.


\section{Results}

    


\begin{table}[t]
    \centering
    \begin{minipage}{0.45\textwidth}
        \centering
    \setlength\tabcolsep{3pt}
    \caption{\RebuttalChange{Performance when running NN-20, NN-50, NN-100. Strix$_{\text{best}}$ denotes the case with highest performance, while Strix$_{\text{128bit}}$ denotes the case when Strix achieves 128 bit security.}}  
    \begin{tabular}{lrrrr}
    \toprule[1pt]
        \textbf{Scheme}& \textbf{Security} & \textbf{NN-20} & \textbf{NN-50} & \textbf{NN-100} \\
        \midrule[0.5pt]
       Baseline-TFHE\cite{zama2022Concrete} & 128-bit   & 64.60s & 129.25s & 263.54s \\
       \RebuttalChange{Strix$_{\text{128bit}}$\cite{putra2023strix}} & 128-bit  & 434.44ms & 1193.77ms & 1511.77ms \\
       \RebuttalChange{Strix$_{\text{best}}$\cite{putra2023strix}} & 80-bit  & 78.96ms & 148.73ms & 551.28ms \\
       \emph{\SolutionName} & 128-bit  & \Tofill{69.86}ms & \Tofill{146.26}ms & \Tofill{277.13}ms \\
       \bottomrule[1pt]
    \end{tabular}
    \label{tab:result:deep-nn-perf}
    \end{minipage}
    \hfill 
    \hfill
    \vspace{10pt}
    \begin{minipage}{0.45\textwidth}
        \centering
    \caption{Performance of Scheme Conversion algorithm (ms). }
    \begin{tabular}{lrrr}
    \toprule[1pt]
    \textbf{Scheme} & \textbf{$\mathbf{n_{slot}}$ = 2} & \textbf{$\mathbf{n_{slot}}$ = 8} & \textbf{$\mathbf{n_{slot}}$ = 32} \\
    \midrule[0.5pt]
    \multicolumn{1}{l}{Baseline-SC}  & \multicolumn{1}{r}{364}  & \multicolumn{1}{r}{492}  & \multicolumn{1}{r}{1,168}                       \\
    \multicolumn{1}{l}{\emph{\SolutionName}} & \multicolumn{1}{r}{\Tofill{0.049}} & \multicolumn{1}{r}{\Tofill{0.063}}  & \multicolumn{1}{r}{\Tofill{0.142}}   \\ 
    \bottomrule[1pt]
    
    \end{tabular}
    \label{tab:result:schemeswitch}
    \end{minipage}
    \vspace{-10pt}
\end{table}





\begin{figure}[t]
  \begin{minipage}[b]{\linewidth}
    \centering
    \includegraphics[width=\linewidth]{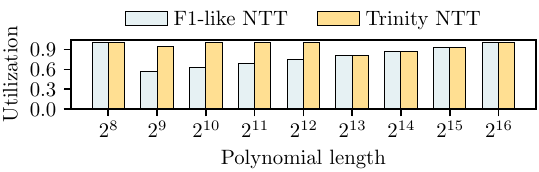}
    \vspace{-15pt}
    \caption{Utilization comparison of NTT unit. }
    \label{fig:result:ntt_util}
  \end{minipage} 
  \begin{minipage}[b]{\linewidth}
    \centering
    \includegraphics[width=0.8\linewidth]{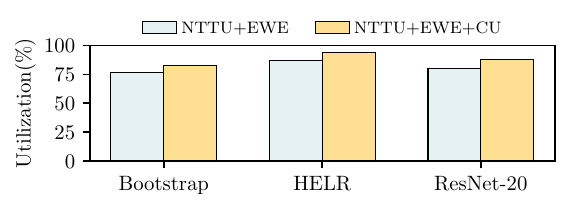}
    \vspace{-10pt}
    \caption{\RebuttalChange{Utilization of NTTU+EWE in {\emph{\SolutionName}-CKKS$_{\text{IP-use-EWE}}$} and NTTU+EWE+CU in \emph{\SolutionName}. }} 
    \label{fig:result:cu_impact_on_ckks-util}
    \vspace{-5pt}
  \end{minipage}
\end{figure}

\begin{table}[t]
    \centering
    \caption{\RebuttalChange{Performance within hybrid-scheme appilications.}}
    \begin{tabular}{crr}
    \toprule[1pt]
    \textbf{Scheme}     & HE3DB-4096 & HE3DB-16384 \\
    
    \midrule[0.5pt]
    Baseline-Hybrid     & 3,012s & 11,835s \\
    SHARP+Morphling     & 5.64s & 22.55s \\
    \emph{\SolutionName}     & 0.42s & 1.68s\\
    \bottomrule[1pt]
    \end{tabular}
    \label{tab:my_label}
    \vspace{-15pt}
\end{table}

\begin{figure}[t] 
  \begin{minipage}[b]{\linewidth}
    \centering
    \includegraphics[width=0.8\linewidth]{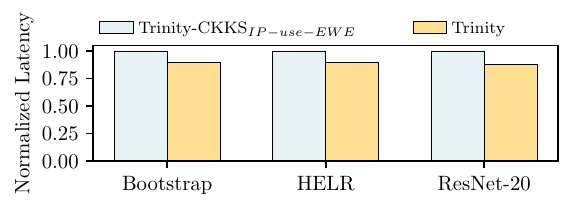}
    \vspace{-5pt}
    \caption{\RebuttalChange{Normalized latency comparison between \emph{\SolutionName}-CKKS$_{\text{IP-use-EWE}}$ and Trinity within CKKS workloads (normalized to \emph{\SolutionName}-CKKS$_{\text{IP-use-EWE}}$). }}
    \label{fig:result:app-perf-for-diff-arch-choice}
  \end{minipage} 
  \begin{minipage}[b]{\linewidth}
    \centering
    \includegraphics[width=0.8\linewidth]{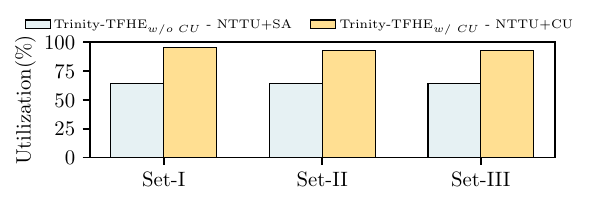}
    \caption{Utilization of \emph{\SolutionName}-TFHE$_{\text{w/o CU}}$ and \emph{\SolutionName}-TFHE$_{\text{w/ CU}}$ when executing PBS. } 
    \label{fig:result:cu_impact_on_pbs-util}
  \end{minipage}
\end{figure}


\begin{table}[t]
    \centering
    \caption{Circuit area and power. }
    \begin{tabular}{l r r}
    \toprule[1pt]
    \textbf{Component} & \textbf{Area(mm$^2$)} & \textbf{Power(W)}\\
    \midrule[0.5pt]
    2$\times$ NTTU     & 3.20 & 4.24   \\
    1$\times$CU-1     & 0.18 & 0.31 \\
    4$\times$CU-2     & 1.44 & 2.48 \\
    1$\times$CU-3     & 0.55 & 0.93 \\
    AutoU     & 0.04 & 0.22 \\
    Rotator     & 2.40 & 8.57 \\
    EWE     & 1.87 & 4.47 \\
    VPU     & 0.05 & 0.07 \\
    \makecell[lt]{NoC (intergroup and\\ intragroup)} & 0.10 & 13.24  \\
    local buffer & 6.45 & 1.41 \\
    \hline
    \textbf{cluster}    & 16.28 & 35.94  \\
    \hline
    4$\times$ cluster     & 65.12 & 143.76   \\
    inter-cluster NoC     & 20.60 & 27.00 \\
    scratchpad     & 41.94 & 26.80 \\
    HBM PHY     & 29.60 & 31.80 \\
    \midrule[0.5pt]
    \textbf{Total} & 157.26 & 229.36 \\
    \bottomrule[1pt]
    \end{tabular}
    \label{tab:result:circuit-area}
\end{table}



\begin{figure}[t]
  \begin{minipage}[b]{\linewidth}
    \centering
    \includegraphics[width=0.8\linewidth]{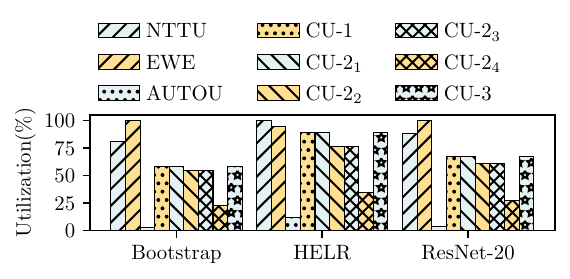}
    \vspace{-10pt}
    \caption{\RebuttalChange{Component utilization within CKKS workloads. }}
    \label{fig:result:ckks_app_util}
      
  \end{minipage}
  \begin{minipage}[b]{\linewidth}
    \centering
    \includegraphics[width=0.8\linewidth]{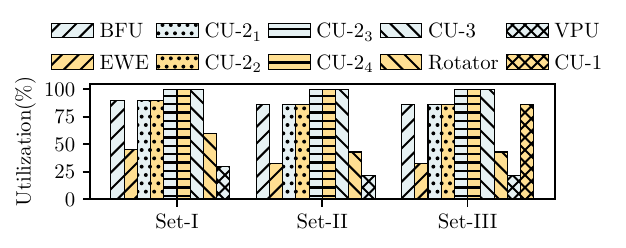}
    \vspace{-10pt}
    \caption{Component utilization within TFHE PBS. }
    \label{fig:result:pbs_util}
    \vspace{10pt}
  \end{minipage}
\end{figure}

\begin{table}[t]\scriptsize
\setlength\tabcolsep{2pt}
\renewcommand{\arraystretch}{1.4}
    \centering
    \caption{Comparison with the state-of-the-art FHE accelerators. BW. denotes bandwidth, Cap. denotes capability.   
    }

    \begin{tabular}{|c|c|c|c|c|c|c}
		\hline
		\multirow{2}{*}{}&{CraterLake\cite{samardzic2022craterlake}}&{{SHARP\cite{kim2023sharp}}}& Morphling\cite{prasetiyo2024morphling}&{\emph{\SolutionName}}  \\ 
        \hline
		\makecell[t]{Scheme \\support}    &{CKKS}&{CKKS}&{TFHE}&{\makecell[tl]{CKKS; TFHE;\\CKKS$\longleftrightarrow$TFHE}}  \\
		\hline
		Word Length    &{28-bit}&{36-bit}&{32-bit}&{36-bit}  \\
        \hline
		Core Freq.    &{1GHz}&{1GHz}&{1.2GHz}&{1GHz}  \\ 
		\makecell[t]{Off-chip \\Mem BW.}    &{1TB/s}&{1TB/s}&{310GB/s}&{1TB/s}  \\ 
        \hline
		\makecell[t]{On-chip \\Mem Cap.}    &{282MB}&{198MB}&\makecell[t]{11MB}&\makecell[t]{191MB}  \\ 
		\makecell[t]{On-chip\\ Mem BW.}    &\makecell[t]{84TB/s}&\makecell[t]{72TB/s}&\makecell[t]{-}&\makecell[t]{36TB/s(SPM);\\135TB/s(local buffer)}  \\ 
		\hline
		Technology	&12$\text{nm}$ & 7$\text{nm}$&{28nm} &{7\text{nm}} \\
		\hline

		Area	&\makecell[t]{472.3$\text{mm}^2$ \\(12nm)}& \makecell[t]{178.8$\text{mm}^2$ \\(7nm)}&\makecell[t]{4$\text{mm}^2$(7nm) \\ 13$\text{mm}^2$(12nm)\\ 74$\text{mm}^2$(28nm)}&\makecell[t]{157.26$\text{mm}^2$(7nm) \\ 462.15$\text{mm}^2$(12nm)}\\
		\hline
  
		Power &320W& - &53.00W&229.36W \\
		\hline
	\end{tabular}
    \label{tab:result:accelerator_comparison}
    \vspace{-10pt}
\end{table}

\begin{figure}[t]
  \begin{minipage}[b]{\linewidth}
    \centering
        \includegraphics[width=0.8\linewidth]{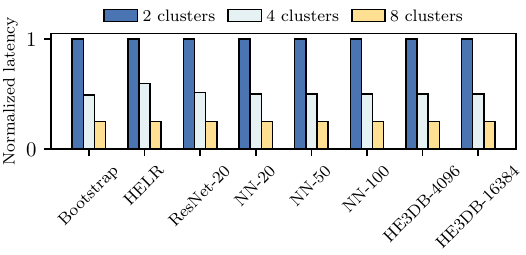}
    \caption{\RebuttalChange{Normalized latency within CKKS, TFHE and hybrid scheme applications under different accelerator configurations. Normalized to 2 clusters. }}
    \label{fig:result:sensi-perf}
  \vspace{-10pt}
  \end{minipage}
\end{figure}

\begin{figure}[t]
    \centering
    \includegraphics[width=1.0\linewidth]{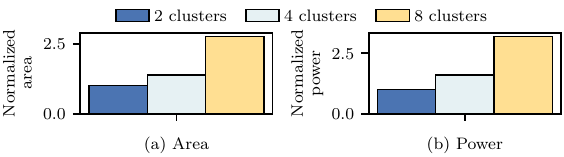}
\caption{\RebuttalChange{Normalized area and power under different accelerator configurations. Normalized to 2 clusters.}}  
\label{fig:result:sensi-area-power}  
\end{figure}

\subsection{Performance}

In this section, we compare the performance of \emph{\SolutionName} against CPUs, GPUs, and state-of-the-art ASIC accelerators for the CKKS, TFHE, and Scheme Conversion tasks. Additionally, we conduct a series of comprehensive analyses to explain the sources of performance benefits.

\RebuttalChange{As shown in Table \ref{tab:result:app-perf}, \emph{\SolutionName} achieves an average performance improvement over SHARP\cite{kim2023sharp} by 1.49$\times$ and up to 1.85$\times$ when executing HELR.} As shown in Table \ref{tab:result:tfhe-pbs-perf}, in terms of PBS throughput, \emph{\SolutionName} outperforms Morphling\cite{prasetiyo2024morphling} by average 4.23$\times$ and up to 4.32$\times$ under Set-III. Additionally, as shown in Table \ref{tab:result:deep-nn-perf}, when computing NN-$x$, \emph{\SolutionName} demonstrates a speedup over Baseline-TFHE by an average 919.3$\times$ and up to 950.9$\times$ within the NN-100 model. \RebuttalChange{Under the same security of 128-bit, compared with Strix, \emph{\SolutionName} achieves a performance improvement of average 6.51$\times$.  Compared with the best case of Strix, \emph{\SolutionName} achieves a performance improvement of average 1.31$\times$ with better security of 128-bit, while the best case of Strix achieves 80-bit security. }When considering Scheme Conversion from TFHE to CKKS, as shown in Table \ref{tab:result:schemeswitch}, \emph{\SolutionName} achieves an average speedup over baseline-SC by 7,814$\times$. \RebuttalChange{For hybrid FHE appilications, \emph{\SolutionName} outperforms baseline-hybrid on average by 7,107$\times$. Compared with the SOTA accelerator, \emph{\SolutionName} outperforms SHARP+Morphling on average by 13.42$\times$.}

\subsection{Hardware efficiency for CKKS and TFHE}

\textbf{The performance impact of CU on CKKS.} 
To validate the effectiveness of our design, we conducted a series of analyses. For CKKS workloads, we first examined the benefits of deploying CUs for NTT computation. As shown in Figure \ref{fig:result:ntt_util}, the NTT design in \emph{\SolutionName} demonstrates an average improvement in utilization by \Tofill{1.2$\times$}, which can be attributed to the flexible mapping capabilities of CUs for NTT computation. Additionally, we analyzed the impact of using CUs for Inner Product computation. As shown in Figure \ref{fig:result:cu_impact_on_ckks-util}, for CKKS workloads, \emph{\SolutionName} shows a significant utilization improvement over \emph{\SolutionName}-CKKS$_{\text{IP-use-EWE}}$ by \Tofill{1.08$\times$}. This enhancement in utilization improves the \emph{\SolutionName}'s performance for CKKS workloads. \RebuttalChange{As shown in Figure \ref{fig:result:app-perf-for-diff-arch-choice}, \emph{\SolutionName} outperforms \emph{\SolutionName}-CKKS$_{\text{IP-use-EWE}}$ by average {1.12$\times$} and up to {1.13$\times$} for {ResNet-20}.}

\textbf{The performance impact of CU on TFHE.} 
As shown in Table \ref{tab:result:tfhe-pbs-perf}, \emph{\SolutionName}-TFHE$_{\text{w/o CU}}$ exhibits an average performance reduction by \Tofill{27.1\%}, whereas \emph{\SolutionName}-TFHE$_{\text{w/ CU}}$ achieves a significant performance improvement over Morphling by average \Tofill{1.27$\times$}. The performance benefit of \emph{\SolutionName}-TFHE$_{\text{w/ CU}}$ can be attributed to the balanced workload enabled by the flexible CU mapping and the reuse of CU in NTT computations. As shown in Figure \ref{fig:result:cu_impact_on_pbs-util}, \emph{\SolutionName}-TFHE$_{\text{w/ CU}}$ demonstrates an average utilization improvement over \emph{\SolutionName}-TFHE${_\text{w/o CU}}$ by \Tofill{1.45$\times$}.  

\textbf{Utilization of other components. }
\RebuttalChange{As shown in Figure \ref{fig:result:ckks_app_util}, \emph{\SolutionName} achieves an average utilization exceeding {48\%} when executing CKKS workloads. }Similarly, Figure \ref{fig:result:pbs_util} shows that under three different parameter sets for TFHE PBS, \emph{\SolutionName} maintains an average utilization above \Tofill{64\%}. The consistently high utilization across both CKKS and TFHE workloads underscores the efficiency of \emph{\SolutionName}, which is largely due to the capability of \emph{\SolutionName} to balance workloads in both schemes.

\subsection{Area and Power}

\textbf{Area. }
Table \ref{tab:result:circuit-area} presents the circuit area of \emph{\SolutionName} by components. Compared to the total area of SHARP and Morphling, \emph{\SolutionName} achieves an area reduction of \Tofill{15\%} and outperforms each individually. This result demonstrates the high computational efficiency of \emph{\SolutionName}, which is attributed to its configurable architecture design.


\textbf{Power. }
As illustrated in Table \ref{tab:result:accelerator_comparison}, in comparison with CraterLake, \emph{\SolutionName} achieves a substantial reduction of \Tofill{28.5\%} in circuit power while substantially outperforming CraterLake. This result indicates that \emph{\SolutionName} has considerable power efficiency.

\subsection{\RebuttalChange{The overhead of supporting both CKKS and TFHE}}

\RebuttalChange{To fully support both CKKS and TFHE, AutoU, EWE, Rotator and VPU are required in the accelerator. These components cannot be utilized in both schemes. Nonetheless, these components constitute only a small proportion of the total area of \emph{\SolutionName}, which is 11.08\%.}

\subsection{\RebuttalChange{Sensitivity study to the number of clusters}}

\RebuttalChange{In this section, we conduct a sensitivity study to the cluster number. As shown in Figure \ref{fig:result:sensi-perf}, when the cluster number increases from 4 to 8, \emph{\SolutionName} achieves a performance improvement of average 2.04$\times$, with an area increment of 2$\times$ (shown in Figure \ref{fig:result:sensi-area-power}(a)). On the contrary, as the cluster number decreases, the end-to-end performance is lower, while \emph{\SolutionName} achieves the reduction of circuit area and power consumption by $28\%$ and $36\%$. Therefore, the user can modify the cluster number to meet the requirements for performance or hardware overhead.}

\section{Related Work}
\textbf{CKKS Accelerators.}
Domain-specific accelerators provide significant performance improvements due to highly customized hardware designs and specific dataflow optimizations for particular scenarios. Consequently, numerous CKKS accelerators have been proposed\cite{samardzic2021f1,samardzic2022craterlake,kim2022bts,kim2022ark,kim2023sharp,yang2023poseidon,agrawal2023fab}. When executing CKKS workloads, these accelerators demonstrate substantial performance enhancements. However, they support only the CKKS scheme and cannot leverage the advantages of both the CKKS and TFHE schemes. Furthermore, most feature fixed designs, which are difficult to adapt to the diverse workload characteristics of the CKKS and TFHE schemes. In contrast, \emph{\SolutionName} features a flexible design with configurable units that can adjust the computational component ratio for different kernels, thereby achieving balanced workloads across both schemes.

\textbf{TFHE Accelerators.}
The TFHE scheme imposes significant computational and memory demands. To address these challenges, several accelerators for TFHE have been developed \cite{jiang2022matcha,putra2023strix,prasetiyo2024morphling,feldmann2023fpt}. These accelerators generally achieve higher throughput compared to conventional CPUs. Nonetheless, the use of FFT, which relies on floating-point arithmetic, adds to the hardware complexity and increases power consumption. Although some implementations utilize FFT based on fixed-point or integer arithmetic, they still cannot mitigate the approximation errors inherent in FFT computations. In contrast, \emph{\SolutionName} utilizes NTT for TFHE computations through minor changes on the schemes, taking the NTT advantage of not inducing additional error.

\textbf{Scheme Conversion Algorithm.}
CKKS schemes offer the benefit of SIMD computation, while TFHE schemes enable arbitrary function evaluation. Consequently, it is natural to seek methods that combine the strengths of both schemes. To this end, various Scheme Conversion algorithms and frameworks have been introduced \cite{al2022openfhe,lu2021pegasus}. These algorithms effectively convert between the CKKS and TFHE schemes, operate in the TFHE domain, and then revert to the CKKS scheme. 

\RebuttalChange{\textbf{Scheme Conversion on FPGA.} Recently, an FPGA accelerator targets efficient CKKS bootstrapping using Scheme Conversion has been proposed\cite{agrawal2024heap}. This accelerator is implemented based on 8 FPGA cards as a distributed parallel system. Nonetheless, {\SolutionName} is the first work to explore the construction of multi-modal FHE scheme and Scheme Conversion support on one specific ASIC accelerator. \emph{\SolutionName} not only provides efficient support for CKKS and TFHE, but also supports Scheme Conversion, enabling the runtime switching between CKKS and TFHE. }

\section{Conclusion}
In this paper, we conduct a thorough analysis of the challenges involved in designing an accelerator compatible with both CKKS and TFHE schemes, including the conversions between them. We then introduce \emph{\SolutionName}, a high-performance fully homomorphic encryption (FHE) accelerator that introduces several optimizations: 1) the consistent application of NTT within both the CKKS and TFHE frameworks to minimize approximation errors and reuse computational components during computations; 2) the configurable units that facilitate flexible data mapping across different cryptographic schemes. {Experimental results indicate that \emph{\SolutionName} markedly outperforms SHARP, the leading CKKS accelerator, with a \RebuttalChange{1.49$\times$} average improvement in performance for CKKS workloads, and exceeds Morphling, the advanced accelerator for TFHE workloads, by achieving a 4.23$\times$ average performance increase for TFHE PBS. }Moreover, \emph{\SolutionName} demonstrates a 15\% reduction in circuit area relative to the combined area of SHARP and Morphling.

\section*{Acknowledgment}

The authors thank Liang Chang and the anonymous reviewers for their valuable comments. This research was supported by the National Key R\&D Program of China (Grant No.2023YFB4503201) and the Strategic Priority Research Program of the Chinese Academy of Sciences (Grant No.XDB44030200). Qian Lou and Dingyuan Cao contributed independently to this work, without support from the aforementioned funding sources.

\bibliographystyle{IEEEtranS}
\bibliography{refs}

\end{document}